\documentclass[aps,prd,onecolumn,superscriptaddress,showpacs]{revtex4}

\def\be{\begin{equation}}
\def\ee{\end{equation}}
\def\ba{\begin{eqnarray}}
\def\ea{\end{eqnarray}}
\def\ra{\rightarrow}

\begin{document}

\title{Reparametrization invariance of $B$ decay amplitudes
and implications for new physics searches in $B$ decays} 

\author{F.\ J.\ Botella}
\affiliation{Centro de F\'{\i}sica Te\'{o}rica de Part\'{\i}culas,
	Instituto Superior T\'{e}cnico,
	P-1049-001 Lisboa, Portugal}
\affiliation{Departament de F\'{\i}sica Te\`{o}rica and IFIC,
        Universitat de Val\`{e}ncia-CSIC,
        E-46100, Burjassot, Spain}
\author{Jo\~{a}o P.\ Silva}
\affiliation{Centro de F\'{\i}sica Te\'{o}rica de Part\'{\i}culas,
	Instituto Superior T\'{e}cnico,
	P-1049-001 Lisboa, Portugal}
\affiliation{Instituto Superior de Engenharia de Lisboa,
	Rua Conselheiro Em\'{\i}dio Navarro,
	1900 Lisboa, Portugal}

\date{\today}

\begin{abstract}
When studying $B$ decays within the Standard Model,
it is customary to use the unitarity of the CKM matrix in
order to write the decay amplitudes in terms of only
two of the three weak phases which appear in the
various diagrams.
Occasionally, 
it is mentioned that those two weak phases can be used 
in order to describe any decay amplitude,
even beyond the Standard Model.
Here we point out that,
when describing a generic decay amplitude,
the two weak phases can be chosen completely at will,
and we study the behavior of the decay amplitudes under changes
in the two weak phases chosen as a basis.
Of course,
physical observables cannot depend on such reparametrizations.
This has an impact in discussions of the SM and in attempts
to parametrize new physics effects in the decay amplitudes.
We illustrate these issues by looking at
$B \rightarrow \psi K_S$ and the isospin
analysis in $B \rightarrow \pi \pi$.
\end{abstract}


\maketitle

\section{\label{sec:intro}Introduction}

In the Standard Model (SM) of electroweak interactions,
CP violation appears through one single irremovable
phase in the Cabibbo-Kobayashi-Maskawa (CKM) matrix \cite{CKM},
making it a rather predictive theory.
The goal of $B$ physics experiments is to exploit this
feature and uncover new physics effects.
Reviews may be found,
for example,
in \cite{BLS,CKMfitter-04,Prague04}.

This program is complicated by the presence of uncertain
hadronic matrix elements relating the quark field operators
utilized in writing down the theory with the hadrons detected by experiment.
In some cases,
such hadronic matrix elements can be removed and one is able to
relate experimental observables with parameters in the original
Lagrangian of the electroweak theory.
In the context of CP violating asymmetries,
the situation is sometimes described by the following statements:
i) ``If the decay amplitude depends on only one weak phase,
then we can relate experiment with a parameter appearing in the
original weak Lagrangian'';
or ii) ``If the decay amplitude is written in terms of
two weak phases, then we cannot relate experiment with a
parameter in the original Lagrangian''.
These statements are \textit{imprecise}!
One of the side-benefits of our analysis is the correction
of these sentences.

In this article,
we clarify what can (and cannot) be said about the weak phases
entering a given decay amplitude:
\begin{itemize}
\item We point out that a given decay amplitude
can be described by \textit{any two weak phases},
$\{\phi_{A1}, \phi_{A2}\}$,
chosen completely at random
(as long as they do not differ by a multiple of 180$^\circ$);
\item We distinguish ``experimental weak phases'' from
``theoretical weak phases'',
explaining that the former can be measured if and only if there
is no direct CP violation;
\item We see what happens when we change the basis utilized
to describe the weak phases from $\{\phi_{A1}, \phi_{A2}\}$ 
into $\{\phi_{A1}^\prime, \phi_{A2}^\prime\}$.
Since physical results cannot change under such a reparametrization,
we refer to this property as ``reparametrization invariance''
of the decay amplitudes;
\item We discuss the impact that reparametrization invariance
has on searches for new physics in $B$ decays;
\item And, we relate our observations with statements scattered
in the literature and applicable in very particular special cases,
explaining their generalizations or limitations.
\end{itemize}

In section~\ref{sec:theorem} we show that 
any two weak phases
can be used to describe
a generic $B$ decay amplitude.
In section~\ref{sec:expVSth} we show that one
can ascertain experimentally whether the decay amplitude
can be parametrized exclusively with a single weak phase
and we illustrate with a few examples that such information
may not be enough to determine a weak phase in the original
electroweak Lagrangian.
In section~\ref{sec:pipi} we see how reparametrization
invariance affects the analysis of 
$B \rightarrow \pi \pi$ decays,
in the SM and in the presence of new physics,
turning to the decays $B \rightarrow \psi K$
in section~\ref{sec:psiK}.
In section~\ref{sec:single} we discuss some general features
of observables determined experimentally to
depend on a single weak phase,
which, moreover, coincides with that predicted in the SM;
we identify the types of new physics which may (or may not) be consistent
with such results.
In section~\ref{sec:conclusions} we present our conclusions.

\section{\label{sec:theorem}Parametrizing the weak phase
content of the decay amplitudes}

\subsection{Model independent analysis}

Let us consider the decay of a $B$ meson into some
specific final state $f$.
For the moment,
$B$ stands for $B^+$, $B_d^0$ or $B_s^0$.
When discussing generic features of the decay amplitudes
without reference to any particular model,
it has become commonplace to parametrize
the decay amplitudes as
\ba
A_f
&=&
M_1 e^{i \phi_{A1}} e^{i \delta_1} + M_2 e^{i \phi_{A2}} e^{i \delta_2},
\label{A_f}
\\
\bar A_{\bar f}
&=&
M_1 e^{- i \phi_{A1}} e^{i \delta_1} + M_2 e^{- i \phi_{A2}} e^{i \delta_2},
\label{Abar_fbar}
\ea
where $\phi_{A1}$ and $\phi_{A2}$ are two CP-odd weak phases;
$M_1$ and $M_2$ are the magnitudes of the corresponding terms;
and $\delta_1$ and $\delta_2$ are the corresponding 
CP-even strong phases.
These expressions apply to the decays of a (neutral or charged)
$B$ meson into the final state $f$ and the charge-conjugated decay,
respectively.
For the decay of a neutral $B$ meson into a CP eigenstate with
CP eigenvalue $\eta_f = \pm 1$,
the RHS of Eq.~(\ref{Abar_fbar}) appears multiplied  by $\eta_f$.

In this context,
it is sometimes mentioned that any third weak phase may be
written in terms of the first two \cite{Helen}.
Indeed,
it is easy to show that the impact of a third weak phase
$\phi_{A3}$ in $A_f$ and $\bar A_{\bar f}$
can be described in terms of $\phi_{A1}$ and $\phi_{A2}$,
as long as there are parameters $a$ and $b$ such that
\ba
e^{i \phi_{A3}} &=& a e^{i \phi_{A1}} + b e^{i \phi_{A2}},
\nonumber\\
e^{-i \phi_{A3}} &=& a e^{-i \phi_{A1}} + b e^{-i \phi_{A2}},
\label{phi3}
\ea
are both satisfied.
The solutions are
\begin{eqnarray}
a &=& \frac{\sin{(\phi_{A3} - \phi_{A2})}}{\sin{(\phi_{A1} - \phi_{A2})}},
\nonumber\\
b &=& \frac{\sin{(\phi_{A3} - \phi_{A1})}}{\sin{(\phi_{A2} - \phi_{A1})}},
\label{aandb}
\end{eqnarray}
which are valid if $\phi_{A1} - \phi_{A2} \neq n \pi$,
with $n$ integer,
meaning that (obviously) the same cannot be done with only one weak phase.

This result can be used to write any amplitude,
with an arbitrary number $N$ of distinct weak phases,
in terms of only two.
Indeed,
\ba
A_f &=& 
\tilde{M}_1 e^{i \phi_{A1}} e^{i \tilde{\delta}_1}
+
\tilde{M}_2 e^{i \phi_{A2}} e^{i \tilde{\delta}_2}
+
\sum_{k=3}^{N} 
\tilde{M}_k e^{i \phi_{Ak}} e^{i \tilde{\delta}_k}
\nonumber\\
&=&
M_1 e^{i \phi_{A1}} e^{i \delta_1}
+
M_2 e^{i \phi_{A2}} e^{i \delta_2},
\label{master}
\ea
if
\ba
M_1 e^{i \delta_1} &=&
\tilde{M}_1 e^{i \tilde{\delta}_1}
+
\sum_{k=3}^{N}
a_k 
\tilde{M}_k  e^{i \tilde{\delta}_k},
\nonumber\\
M_2 e^{i \delta_2} &=&
\tilde{M}_2 e^{i \tilde{\delta}_2}
+
\sum_{k=3}^{N}
b_k 
\tilde{M}_k  e^{i \tilde{\delta}_k},
\ea
and
\begin{eqnarray}
a_k &=& \frac{\sin{(\phi_{Ak} - \phi_{A2})}}{\sin{(\phi_{A1} - \phi_{A2})}},
\nonumber\\
b_k &=& \frac{\sin{(\phi_{Ak} - \phi_{A1})}}{\sin{(\phi_{A2} - \phi_{A1})}}.
\label{ak_bk}
\end{eqnarray}

Two questions now arises.
First question: which two weak phases do we take as
our basis $\{\phi_{A1}, \phi_{A2}\}$?
As we shall recall below,
for each decay amplitude there are three choices
which appear natural within the SM.
But the derivation leading to Eq.~(\ref{master}) made no explicit
reference to the weak phases $\{\phi_{A1}, \phi_{A2}\}$ chosen;
it didn't even refer to any particular model for the weak interactions.
It is true that,
within some particular model,
we may look at its Lagrangian for inspiration.
But we \textit{need not} do that.
We may choose for our basis any pair of weak phases 
(as long as they do not differ by a multiple of $180^\circ$);
say, $\{0^\circ, 90^\circ\}$,
or even $\{5^\circ, 10^\circ\}$.

Second question:
what happens when we describe the decay amplitudes
with different sets of weak phases
$\{\phi_{A1}, \phi_{A2}\}$ as our basis?
Consider a second set of weak phases
$\{\phi_{A1}^\prime, \phi_{A2}^\prime\}$.
Using Eqs.~(\ref{phi3}) and (\ref{aandb}),
it is easy to show that
\ba
A_f &=& 
M_1 e^{i \phi_{A1}} e^{i \delta_1}
+
M_2 e^{i \phi_{A2}} e^{i \delta_2},
\nonumber\\
&=&
M_1^\prime e^{i \phi_{A1}^\prime} e^{i \delta_1^\prime}
+
M_2^\prime e^{i \phi_{A2}^\prime} e^{i \delta_2^\prime},
\label{two_conventions}
\ea
as long as
\ba
M_1^\prime e^{i \delta_1^\prime}
&=&
M_1 e^{i \delta_1}
\frac{\sin{(\phi_{A1} - \phi_{A2}^\prime)}}{
\sin{(\phi_{A1}^\prime - \phi_{A2}^\prime)}}
+
M_2 e^{i \delta_2}
\frac{\sin{(\phi_{A2} - \phi_{A2}^\prime)}}{
\sin{(\phi_{A1}^\prime - \phi_{A2}^\prime)}},
\nonumber\\
M_2^\prime e^{i \delta_2^\prime}
&=&
M_1 e^{i \delta_1}
\frac{\sin{(\phi_{A1} - \phi_{A1}^\prime)}}{
\sin{(\phi_{A2}^\prime - \phi_{A1}^\prime)}}
+
M_2 e^{i \delta_2}
\frac{\sin{(\phi_{A2} - \phi_{A1}^\prime)}}{
\sin{(\phi_{A2}^\prime - \phi_{A1}^\prime)}}.
\label{relate}
\ea
Eqs.~(\ref{relate}) tell us how to relate the parameters needed
to describe the decay amplitudes with two different choices
for the pair of weak phases used as a basis \cite{details}.
We stress that these weak phases may be chosen completely
at will. Any set will do.

Of course,
physical results cannot change under such a reparametrization;
we refer to this property as ``reparametrization invariance''.
But,
this property implies that we must be careful with our wording and
interpretations,
especially when discussing new physics effects.
This is strikingly clear in those situations usually
described as depending on a single weak phase,
to be studied in section~\ref{sec:single}.

\subsection{Remarks on the Standard Model}

We can use the results of the previous section
in order to place in a more general context some
statements commonly made about the SM.
In the SM, the $\bar b \rightarrow \bar q$ transitions ($q=d,s$)
involve the three CKM structures $V_{ub}^\ast V_{uq}$,
$V_{cb}^\ast V_{cq}$,
and $V_{tb}^\ast V_{tq}$.
A generic decay amplitude may be written as
\be
A(\bar b \rightarrow \bar q)
=
V_{ub}^\ast V_{uq} A_u
+ V_{cb}^\ast V_{cq} A_c
+ V_{tb}^\ast V_{tq} A_t,
\label{decay}
\ee
where the $A_i$ ($i=u,c,t$) involve the relevant hadronic matrix elements
with the corresponding CP-even strong phases.
But,
the unitarity of the CKM matrix,
\be
V_{ub}^\ast V_{uq} + V_{cb}^\ast V_{cq} + V_{tb}^\ast V_{tq} = 0,
\label{unitarity}
\ee
can be used to express everything in terms of only two weak phases.
Of course,
there are three such possibilities:
\ba
A(\bar b \rightarrow \bar q)
&=&
V_{ub}^\ast V_{uq} \left( A_u - A_t \right)
+ V_{cb}^\ast V_{cq} \left( A_c - A_t \right),
\label{uc}
\\
&=&
V_{ub}^\ast V_{uq} \left( A_u - A_c \right)
+ V_{tb}^\ast V_{tq} \left( A_t - A_c \right),
\label{ut}
\\
&=&
V_{cb}^\ast V_{cq} \left( A_c - A_u \right)
+ V_{tb}^\ast V_{tq} \left( A_t - A_u \right).
\label{ct}
\ea
This strategy is followed universally.
In the context of $B_d \rightarrow \pi^+ \pi^-$ decays,
the parametrization in Eq.~(\ref{uc}) is known as the ``$c$-convention'' and
the parametrization in Eq.~(\ref{ut}) is known as the ``$t$-convention'';
the relation among them has been discussed in detail
by Gronau and Rosner in \cite{GR-ct}.
We will name the parametrization in Eq.~(\ref{ct}) the 
``$p$-convention'' (for penguin),
as it does not contain the CKM structure usually associated
with the tree level diagram.
In the SM,
the amplitudes $A_i$ could be calculated exactly if we
knew how to calculate the corresponding hadronic matrix
elements \cite{gauge}.

Therefore,
in the context of the SM,
Eqs.~(\ref{uc}) through (\ref{ct})
provide us with three natural choices for the pair of
weak phases $\{\phi_{A1}, \phi_{A2}\}$ chosen as the basis
for Eq.~(\ref{master}).
For example,
it the $t$-convention of Eq.~(\ref{ut}),
we would take
\ba
M_1 e^{i \phi_{A1}} e^{i \delta_1}
&=&
V_{ub}^\ast V_{uq} \left( A_u - A_c \right),
\nonumber\\
M_2 e^{i \phi_{A2}} e^{i \delta_2}
&=&
V_{tb}^\ast V_{tq} \left( A_t - A_c \right).
\ea
But,
although the unitarity of the CKM matrix was utilized in reaching
Eqs.~(\ref{uc})--(\ref{ct}),
unitarity is not needed in order to justify any of these basis choices.
For example,
we can use the weak phases in $V_{ub}^\ast V_{uq}$ and
$V_{tb}^\ast V_{tq}$ as a basis,
regardless of whether the CKM is unitary or not.
Moreover,
although these (three) choices are natural and useful within the SM,
they are not mandatory;
not even within the SM.
We stress our main point:
one may choose for the basis any pair of weak phases 
(as long as they do not differ by a multiple of $180^\circ$);
say, $\{0^\circ, 90^\circ\}$,
or even $\{5^\circ, 10^\circ\}$.

\section{\label{sec:expVSth}''Experimental'' weak phases
versus ``theory'' weak phases}

\subsection{\label{subsec:experimental}``Experimental'' determination of the
presence of a single weak phase}

It turns out that one can determine experimentally,
at least in principle,
if the decay amplitude of a neutral $B$ meson can (or not)
be written in terms of a single weak phase.
In order to clarify this statement,
we recall that the full description of neutral meson decays involves 
Eqs.~(\ref{A_f}),
(\ref{Abar_fbar}),
and also the mixing parameter
\be
\frac{q_B}{p_B}
=
e^{2i\phi_M},
\label{q/p_parametrization}
\ee
in the combination
\ba
\lambda_f 
&=& 
\frac{q_B}{p_B}
\frac{\bar A_f}{A_f}
\\
&=&
\eta_f e^{-2i \phi_1}
\frac{1+r e^{i(\phi_1 - \phi_2)} e^{i \delta}}{
1+r e^{- i(\phi_1 - \phi_2)} e^{i \delta}}\ ,
\label{Lf_r_and_delta}
\ea
where
$\phi_1 \equiv \phi_{A1}-\phi_M$,
$\phi_2 \equiv \phi_{A2}-\phi_M$,
$\delta=\delta_2-\delta_1$,
and $r=M_2/M_1$.
We have assumed that $|q_B/p_B|=1$,
meaning that the CP violation in $B - \overline{B}$ mixing is
negligible.
To set the notation,
we recall 
in the appendix
that $\lambda_f$
is measurable from the decay rates through
\ba
S_f &\equiv&
\frac{2 \mbox{Im}( \lambda_f )}{1 + |\lambda_f|^2}
=
- \eta_f 
\frac{\sin{(2 \phi_1)} + 2 r \sin{(\phi_1 + \phi_2)} \cos{\delta} + 
r^2 \sin{(2 \phi_2)}}{
1 + 2 r \cos{(\phi_1 - \phi_2)} \cos{\delta} + r^2},
\label{S_f_detail}
\\
C_f &\equiv&
\frac{1 - |\lambda_f|^2}{1 + |\lambda_f|^2}
=
\frac{2 r \sin{(\phi_1 - \phi_2)} \sin{\delta}}{
1 + 2 r \cos{(\phi_1 - \phi_2)} \cos{\delta} + r^2},
\label{C_f_detail}
\ea
since
\begin{equation}
\lambda_f =
\frac{1}{1+C_f} \left( 
\pm \sqrt{1 - C_f^2 - S_f^2} + i S_f \right)
\label{Lf_from_CS}
\end{equation}
For simplicity,
we will assume in the following that $S_f$ and $C_f$
can be measured with absolute precision.

We now claim that 
\textit{$C_f = 0$ if and only if the decay amplitude
is dominated by a single weak phase}.
Moreover,
\textit{in such cases $S_f$ determines that weak phase},
up to discrete ambiguities.
It is clear that a decay dominated by a single weak phase
leads to $C_f=0$,
so we only have to show the converse.
Let us assume that $C_f=0$.
From Eq.~(\ref{C_f_detail}),
this implies that:
i) $r=0$ (and there is only one amplitude/weak phase);
or that ii) $\phi_1=\phi_2$ (and there is only one weak phase);
or that iii) $\delta_1=\delta_2$ \cite{pi}.
In the last case, we can always find a
magnitude $M_3$ and a weak phase $\phi_{A3}$ such that
\ba
A_f
&=&
\left( M_1 e^{i \phi_{A1}} + M_2 e^{i \phi_{A2}} \right) e^{i \delta_1}
=
M_3 e^{i \phi_{A3}} e^{i \delta_1},
\label{A_f_to3}
\\
\bar A_{\bar f}
&=&
\eta_f
\left( M_1 e^{- i \phi_{A1}} + M_2 e^{- i \phi_{A2}} \right) e^{i \delta_1}
=
\eta_f
M_3 e^{- i \phi_{A3}} e^{i \delta_1}.
\label{Abar_fbar_to3}
\ea
These equalities are satisfied by
\ba
M_3^2
&=&
M_1^2 + M_2^2 + 2 M_1 M_2 \cos{(\phi_1 - \phi_2)},
\label{sol_M3}
\\
e^{- 2 i \phi_{A3}}
&=&
\frac{M_1 e^{- i \phi_{A1}} + M_2 e^{- i \phi_{A2}}}{
M_1 e^{i \phi_{A1}} + M_2 e^{i \phi_{A2}}}
=
\frac{ 1 + r e^{i (\phi_{A1} - \phi_{A2})}}{
1 + r e^{- i (\phi_{A1} - \phi_{A2})}}
e^{- 2 i \phi_{A1}},
\label{sol_phi3}
\ea
from which we can always determine $\phi_{A3}$,
because the numerator and the denominator on the RHS of Eq.~(\ref{sol_phi3})
are complex conjugate.
(Of course, the same will not hold if $\delta_1 \neq \delta_2$.)
In cases i) and ii) $S_f = - \eta_f \sin{(2 \phi_1)}$;
in case iii) $S_f = - \eta_f \sin{(2 \phi_3)}$,
where $\phi_3 = \phi_{A3} - \phi_M$.
This completes our proof.

We stress the significance of our result:
if $C_f=0$ then we are sure that the amplitudes may be written
in terms of only one weak phase,
which, moreover,
is measured through $S_f$.
This occurs even for case iii) which was originally written
as containing two distinct weak phases.
Our result seems to contradict the usual simplified
statement that
``if the decay amplitude is determined by only one weak phase,
then we can relate experiment with theory;
if more than one weak phase is involved,
then we cannot''.
The subtle, yet crucial, point is not whether we may write
the decay amplitudes in terms of only one weak phase
(a fact we have just shown can be ascertained experimentally)
but, rather, whether we may write the decay amplitudes
in terms of only one weak phase \textit{which we can identify from
the theoretical Lagrangian}.

\subsection{\label{subsec:theory}Theory faces experiment}

\subsubsection{\label{subsubsec:theory1}Two ``theory'' weak
phases can look like one ``experimental'' weak phase}

Let us consider some decay,
such as $B_d \rightarrow \pi \pi$ in the SM,
which receives contributions from a tree and a penguin
diagram with different weak phases,
$\phi_{A1}$ and $\phi_{A2}$,
respectively,
in some chosen phase convention.
Now imagine that,
by some accident,
their relative strong phase vanishes. 
(Of course, this is a very simplified picture,
but it will illustrate our point.)
In that case,
Eqs.~(\ref{A_f_to3})--(\ref{sol_phi3})
guarantee that we may rewrite the corresponding decay
amplitude as depending on only one weak phase, $\phi_{A3}$.
The problem is that we cannot turn that ``experimental''
information into knowledge about the weak phases
$\phi_1$ and $\phi_2$ which appeared in our
``theoretical'' Lagrangian.
This can be seen clearly in Eq.~(\ref{sol_phi3}):
even if we knew $\phi_1$ from elsewhere,
we would still require knowledge of $r$ in order
to extract $\phi_2$ from the ``experimental''
determination of $\phi_3$.
And,
unfortunately,
$r=M_2/M_1$ depends on the ratio of uncertain
hadronic matrix elements.

In this case,
although we know from experiment that the decay amplitude may
be rewritten in terms of a single weak phase,
(without full knowledge of $r$)
we have no way of turning that information into a determination
on the weak phases present in the theoretical Lagrangian.

\subsubsection{\label{subsubsec:theory2}One ``theory'' weak
phase can be written as two ``theory'' weak phases}

Let us now consider some decay amplitude usually described by a single
weak phase $\phi_{A3}$,
such as the decay amplitude for $B_d \rightarrow \psi K_S$
within the SM.
Clearly,
Eq.~(\ref{phi3}) implies that we could have chosen to describe
the same amplitude with two weak phases which,
as we stress in this article,
could be chosen completely at will
(we will stop mentioning that the two weak phases chosen cannot
differ by a multiple of $180^\circ$).
For example,
we could use Eq.~(\ref{phi3}) in order to rewrite
the single weak phase in terms of,
say,
$\phi_{A1}=5^\circ$ and $\phi_{A2}=10^\circ$.
Using Eqs.~(\ref{phi3}) and (\ref{aandb}) we find
\begin{eqnarray}
A_f =
M_3 e^{i \phi_{A3}} e^{i \delta_{3}}
&=&
\frac{M_3 e^{i \delta_{3}}}{\sin{5^\circ}}
\left[
- \sin{(\phi_{A3} - 10^\circ)}\,
e^{i\, {5^\circ}}
+
\sin{(\phi_{A3} - 5^\circ)}\,
e^{i\, {10^\circ}}
\right],
\nonumber\\
\bar A_f =
M_3 e^{-i \phi_{A3}} e^{i \delta_{3}}
&=&
\frac{M_3 e^{i \delta_{3}}}{\sin{5^\circ}}
\left[
- \sin{(\phi_{A3} - 10^\circ)}\,
e^{-i\, {5^\circ}}
+
\sin{(\phi_{A3} - 5^\circ)}\,
e^{-i\, {10^\circ}}
\right].
\label{nasty}
\end{eqnarray}
In this case,
the amplitudes corresponding to those weak phases must obey
\be
r e^{i \delta}
=
\frac{M_2}{M_1} e^{i (\delta_2 - \delta_1)}
=
- \frac{\sin{(\phi_{A3} - 5^\circ)}}{
\sin{(\phi_{A3} - {10}^\circ)}},
\label{rdelta_12}
\ee
from which $\delta$ is either $0$ or $\pi$ and we get back
the case iii) discussed above.
Again,
the key point is whether we can relate the weak phase
measured by experiment with a parameter in the theoretical
Lagrangian.
In this case we assumed from the start that we could;
we assumed that we could relate $\phi_{A3}$ with a
weak phase in the theoretical Lagrangian.
Rewriting things in terms of two weak phases,
as in Eq.~(\ref{nasty}),
was just a nasty and unneeded complication.

One might worry whether,
when written in terms of the two weak phases
$\{ \phi_{A1}, \phi_{A2} \} = \{ 5^\circ, 10^\circ \}$,
the experimental observables $S_f$ and $C_f$ remain consistent
with the presence of a single weak phase.
Specially given how complicated 
Eqs.~(\ref{S_f_detail}) and (\ref{C_f_detail}) become.
However,
going back to Eqs.~(\ref{S_f_detail}) and (\ref{C_f_detail}),
and substituting $\phi_1 = 5^\circ - \phi_M$,
$\phi_2 = 10^\circ - \phi_M$,
and $r e^{i \delta}$ by Eq.~(\ref{rdelta_12}),
we do indeed recover
$S_f = - \eta_f \sin{(2 \phi_3)}$
and $C_f = 0$.

\subsection{\label{sec:which}Which weak phases can we measure?}

The most striking result of the previous arguments is the following:
\textit{a priori, the weak phases appearing in the parametrization
of the decay amplitudes have no physical meaning.}
Said otherwise,
the fact that some weak phase is written in the decomposition
of $A_f$ does not, by itself, guarantee that that weak phase
will be observable.
This is the only possible conclusion from the fact that we can
\textit{choose at will}
the two weak phases $\{ \phi_{A1}, \phi_{A2} \}$ for our basis,
\textit{c.f.\/} Eqs.~(\ref{two_conventions}),
and that there are infinite such choices.

Although we have already proved this result,
it is interesting to revisit it in the following way.
Let us compare the two basis
$\{ \phi_{A1}, \phi_{A2} \}$ and
$\{ \phi_{A1}^\prime, \phi_{A2} \}$,
with $\phi_{A1} \neq \phi_{A1}^\prime$.
Imagine that there were an algorithm allowing us to
write $\phi_{A1}$ as a function of physical observables.
Then,
given the similarity of the functional forms in
Eqs.~(\ref{two_conventions}),
we would be able to extract $\phi_{A1}^\prime$ with exactly the same
function of the physical observables.
But that would lead to $\phi_{A1} = \phi_{A1}^\prime$,
which contradicts our assumption.
A very particular version of this argument has
been used by London, Sinha and Sinha in \cite{LSS_ct},
when comparing the $c$- and $t$-conventions in
$B \rightarrow \pi \pi$ decays.
Here we have shown that this result is true in complete
generality;
it affects whatever weak phase we use in the decay amplitude,
because we can use any phase $\phi_{A1}$ in the parametrization of
any decay amplitude (as long as we use also a second weak phase
$\phi_{A2} \neq \phi_{A1} + n \pi$).
What we cannot do is parametrize a \textit{generic} decay amplitude
using \textit{exclusively} the weak phase $\phi_{A1}$.

Thus far, we have shown the following:
\begin{itemize}
\item Experimentally,
if $C_f=0$ (corresponding to $|\lambda_f|=1$),
then one can write the decay amplitude in terms of only
one weak phase,
and that phase is measurable through $S_f$.
\item Theoretically,
one usually writes that decay amplitude in terms of the weak phases
appearing in the electroweak Lagrangian.
\item If the theoretical description of the decay amplitude involves
only one weak phase from the electroweak Lagrangian, then
we can identify that phase with the phase measured experimentally.
\item Otherwise,
we have an ``experimental'' weak phase which we cannot turn into
a determination of the ``theory'' weak phases used in the
decomposition of the decay amplitude.
\end{itemize}
These issues are easier to understand by considering a specific
decay within the SM.

\section{\label{sec:pipi}Reparametrization invariance
in $B \rightarrow \pi\pi$ decays}

\subsection{\label{subsec:decay_amp_pipi}Decay amplitudes in
$B \rightarrow \pi \pi$}

In the SM,
the $\bar b \rightarrow \bar d$ decays leading to
$B \rightarrow \pi \pi$ can get contributions
proportional to any of the CKM structures in
Eq.~(\ref{decay}):
\ba
V_{ub}^\ast V_{ud}
&\sim&
A \lambda^3 R_b e^{i \gamma},
\label{btod_phase_1}
\\
V_{cb}^\ast V_{cd}
&\sim&
- A \lambda^3,
\label{btod_phase_2}
\\
V_{tb}^\ast V_{td} 
&\sim&
A \lambda^3 R_t e^{-i \beta}.
\label{btod_phase_3}
\ea
The quantities appearing on the RHS are defined in the appendix
in a rephasing invariant way,
but a convenient phase convention has been used in equating the
RHS to the LHS.
Substitution of Eqs.~(\ref{btod_phase_1})-(\ref{btod_phase_3})
in Eq.~(\ref{unitarity})
leads to the usual form of the unitarity triangle
\be
R_b e^{i \gamma} + R_t e^{-i \beta} = 1.
\label{usual_unit_triangle}
\ee
Thus, 
if inspired by the weak phases which appear in the SM description
of $B \rightarrow \pi \pi$ decays,
we are lead to choosing two among the weak phases
$\gamma$, $\pi$, and $- \beta$.
In the $c$-convention of Eq.~(\ref{uc}),
the phases are $\gamma$ and $\pi$;
in the $t$-convention of Eq.~(\ref{ut}),
the phases are $\gamma$ and $- \beta$;
and, in the $p$-convention of Eq.~(\ref{ct}),
the phases are $\pi$ and $- \beta$.

Of course,
due to reparametrization invariance,
we may equally well choose any other set of two phases as our basis;
we need not look for the SM for inspiration.
For example,
we could again choose to write the decay amplitudes in terms of
$\{ \phi_{A1}, \phi_{A2} \} = \{ 5^\circ, 10^\circ \}$.
Any two different sets are related to each other as in
Eqs.~(\ref{relate}).

Let us apply this relation to the two sets
$\{ \phi_{A1}, \phi_{A2} \} = \{ \gamma, \pi \}$,
and
$\{ \phi_{A1}^\prime, \phi_{A2}^\prime \} = \{ \gamma, - \beta \}$.
Direct application of Eqs.~(\ref{relate}) leads to
\ba
T_t e^{i \delta^T_t}
&=&
T_c e^{i \delta^T_c}
-
\frac{\sin{\beta}}{\sin{(\beta + \gamma)}}
P_c e^{i \delta^P_c},
\nonumber\\
P_t e^{i \delta^P_t}
&=&
-
\frac{\sin{\gamma}}{\sin{(\beta + \gamma)}}
P_c e^{i \delta^P_c},
\label{reproduce}
\ea
where we have used the conventional notation 
$M_1 = T_c$,
$M_2 = P_c$,
$M_1^\prime = T_t$,
and $M_2^\prime = P_t$ \cite{relate_notations}.
Moreover,
$\delta_1 = \delta_c^T$,
$\delta_1^\prime = \delta_t^T$,
$\delta_2^\prime = \delta_t^P$,
and our $\delta_2 = \delta_c^P$
has the opposite sign of that used by Gronau and Rosner in \cite{GR-ct},
because we define the weak phase of the penguin contribution
in the $c$-convention as $\pi$,
rather than $0$ as done there.
Eqs.~(\ref{reproduce}) reproduce Eqs.~(6) through
(10) of reference \cite{GR-ct}.
However,
here it becomes clear that those relations have absolutely nothing to
do with the unitarity of the CKM matrix,
which was clearly not used in deriving Eqs.~(\ref{relate}).
Eqs.~(\ref{reproduce}) result merely from our freedom to reparametrize the
decay amplitudes.

In some sense,
the relations in Eqs.~(\ref{reproduce}) do not even 
bear any relation to the SM.
Because this might be difficult to accept,
let us explain it in detail.
Imagine that there is no new physics contribution 
affecting the determinations of
$|V_{ub}|$,
$\Delta m_d$,
and the CP violating asymmetry in $B_d \rightarrow \psi K_S$.
These measurements determine the Wolfenstein parameters $\eta$ and $\rho$,
and, thus, the phases $\beta$ and $\gamma$,
albeit with errors.
Let us now assume that there is a substantial
new physics contribution to $B_d \rightarrow \pi^+ \pi^-$.
Then,
we may still parametrize the decay amplitude in terms
of the known phases
$\{ \phi_{A1}, \phi_{A2} \} = \{ \gamma, \pi \}$,
or, alternatively,
in terms of the known phases
$\{ \phi_{A1}^\prime, \phi_{A2}^\prime \} = \{ \gamma, - \beta \}$.
And Eqs.~(\ref{reproduce}) still tell us how to go from one parametrization
to the next.
Nevertheless,
there is a difference between
the unreal scenario we are considering here and the SM.
In the unreal scenario considered here we have absolutely no
access to the magnitudes
$T_t$, $P_t$, $T_c$, and $P_c$.
In contrast,
in the SM we might, in principle,
calculate these magnitudes within some procedure,
such as QCD factorization \cite{QCDF} or perturbative
QCD \cite{PQCD}.

An exercise similar to the one leading to the relations
in Eqs.~(\ref{reproduce})
allows us to relate the quantities in the
$p$-convention with the others.
We find
\ba
M_{1p} e^{i \delta_{1p}}
&=&
- \frac{\sin{(\beta + \gamma)}}{\sin{\beta}}
T_c e^{i \delta^T_c}
+
P_c e^{i \delta^P_c},
\nonumber\\
M_{2p} e^{i \delta_{2p}}
&=&
- \frac{\sin{\gamma}}{\sin{\beta}}
T_c e^{i \delta^T_c},
\label{reproduce-p-c}
\ea
and
\ba
M_{1p} e^{i \delta_{1p}}
&=&
- \frac{\sin{(\beta + \gamma)}}{\sin{\beta}}
T_t e^{i \delta^T_t},
\nonumber\\
M_{2p} e^{i \delta_{2p}}
&=&
- \frac{\sin{\gamma}}{\sin{\beta}}
T_t e^{i \delta^T_t}
+
P_t e^{i \delta^P_t},
\label{reproduce-p-t}
\ea
where we have denoted the magnitudes and strong phases in the
$p$-convention by $M_{1p}$, $M_{2p}$, $\delta_{1p}$,
and $\delta_{2p}$ \cite{relate_notations}.

%

\subsection{\label{subsec:neutral_pipi}Parameter counting
for the CP asymmetry in $B_d \rightarrow \pi^+ \pi^-$}

Let us go back to the $c$-convention,
\be
A_{\pi^+ \pi^-}
=
T_c\, e^{i \delta^T_c} e^{i \gamma}
+
P_c\, e^{i \delta^P_c} e^{i \pi},
\label{A_pipi_c}
\ee
and use the new notation $\phi_M = - \tilde{\beta}$ for the
weak phase in mixing.
From Eq.~(\ref{Lf_r_and_delta}),
we find
\be
\lambda_{\pi^+ \pi^-}
=
e^{- 2 i \tilde{\beta}}\,
\frac{e^{- i \gamma} + z_c}{
e^{i \gamma} + z_c},
\label{lambda_in_c-convention}
\ee
where
\be
z_c = e^{i \pi}\, \frac{P_c\, e^{i \delta^P_c}}{
T_c\, e^{i \delta^T_c}}.
\ee
Notice that this parametrization is completely general.
Any new physics model (with $|q_B/p_B|=1$) can be brought to
this form.
If the new physics affects the phase of the mixing but does not affect
substantially the $\bar b \rightarrow \bar c c \bar s$ decay
amplitudes,
then $\tilde{\beta}$ is the phase measured in the decays
$B_d \rightarrow \psi K$.
In that case,
the two measurements contained in $\lambda_{\pi^+ \pi^-}$
(\textit{i.e.}, its magnitude and phase, or, alternatively,
$S_{\pi^+ \pi^-}$ and $C_{\pi^+ \pi^-}$)
depend on three parameters:
$\gamma$, the phase and the magnitude of $z_c$.

Similarly,
in the $t$-convention
\be
A_{\pi^+ \pi^-}
=
T_t\, e^{i \delta^T_t} e^{i \gamma}
+
P_t\, e^{i \delta^P_t} e^{- i \beta},
\label{A_pipi_t}
\ee
and 
\be
\lambda_{\pi^+ \pi^-}
=
e^{- 2 i \tilde{\beta}}\,
\frac{e^{- i \gamma} + z_t\, e^{i \beta}}{
e^{i \gamma} + z_t\, e^{-i \beta}}
=
e^{- 2 i (\tilde{\beta} - \beta)}\,
\frac{e^{- i (\beta + \gamma)} + z_t}{
e^{i (\beta + \gamma)} + z_t},
\label{lambda_in_t-convention}
\ee
where
\be
z_t = \frac{P_t\, e^{i \delta^P_t}}{
T_t\, e^{i \delta^T_t}}.
\ee
In this notation,
assuming that $\tilde{\beta}$ has been measured,
the two observables depend on four parameters:
$\gamma$, $\beta$, the phase and magnitude of $z_t$.

In the SM,
$\tilde{\beta} \equiv \beta$ and the parameter
counting in the two notations coincides.
But,
in the presence of new physics in the mixing,
the parameter counting does not coincide;
the $t$-convention seems to have one more unknown.
Why does this occur?
We have assumed that $\tilde{\beta}$ is measured
in $B_d \rightarrow \psi K$ decays.
So,
we only have as many unknowns as those present in
$\bar A_f/A_f$.
When the decay amplitudes are written in terms of
only two weak phases,
there are four unknowns:
the two weak phases $\{\phi_{A1}, \phi_{A2}\}$;
the ratio of magnitudes;
and the difference of strong phases.
In the $c$-convention,
we choose for the first phase the theory variable $\gamma$,
and for the second phase the constant phase $\pi$.
In contrast,
in the $t$-convention,
we choose for the first phase the theory variable $\gamma$,
and for the second phase another theory variable $- \beta$.
So,
natural as they may be,
the different choices lead to different parameter
counting.

Using Eqs.~(\ref{reproduce}) we find the relation between the
two hadronic parameters to be
\be
z_c = \frac{z_t \sin{(\beta + \gamma)}}{
\sin{\gamma} - z_t \sin{\beta}}.
\label{magic1}
\ee
So, in a way,
the three parameters which appeared in the $t$-convention as
$z_t$ and $\beta$ are,
in the $c$-convention,
reorganized into the two parameters in $z_c$.
It is also interesting to understand why the extra weak phase $\beta$
goes completely unnoticed when one stays within the SM
and calculates (with some prescription) the hadronic matrix elements
in the two conventions.
The reason lies in the unitarity of the CKM matrix,
due to which
\ba
R_b &=& \frac{\sin{\beta}}{\sin{(\beta+\gamma})},
\nonumber\\
R_t &=& \frac{\sin{\gamma}}{\sin{(\beta+\gamma})}.
\ea
As a result,
\be
z_c = \frac{z_t}{R_t - z_t\, R_b}.
\label{magic2}
\ee
Eq.~(\ref{magic1}) seemed to involve the weak phases.
However,
due to CKM unitarity,
this dependence on the weak phases is hidden
as CP-conserving quantities $R_t$ and $R_b$
in the relation between the hadronic parameters
in the two conventions ($z_c$ and $z_t$).
%
This reflects the fact that the model calculations of $z_c$ depend
on the CP-conserving constraints on the CKM matrix.
Thus,
any knowledge about $\gamma$
obtained by combining the $c$-convention in
Eq.~(\ref{lambda_in_c-convention})
with some model calculation of $z_c$ provides a consistency check
within the SM, but it may not yield an independent determination of
the phase $\gamma$ of the generalized CKM matrix,
valid in models with non-unitary CKM matrices.

As a further example,
we write
\begin{eqnarray}
A_{\pi^+ \pi^-}
&=&
A_5\, e^{i \delta_5} e^{i\; 5^\circ}
+
A_{10}\, e^{i \delta_{10}} e^{i\; 10^\circ},
\nonumber\\
\bar A_{\pi^+ \pi^-}
&=&
A_5\, e^{i \delta_5} e^{- i\; 5^\circ}
+
A_{10}\, e^{i \delta_{10}} e^{- i\; 10^\circ},
\end{eqnarray}
and we use Eqs.~(\ref{relate}) in order to relate
the basis set $\{ \phi_{A1}, \phi_{A2} \} = \{ \gamma, \pi \}$,
with
$\{ \phi_{A1}^\prime, \phi_{A2}^\prime \} = \{ 5^\circ, 10^\circ \}$.
We find
\ba
- A_5\, e^{i \delta_5}
&=&
\frac{\sin{(\gamma - 10^\circ)}}{\sin{(5^\circ)}}
T_c\, e^{i \delta^T_c}
+
2 \cos{(5^\circ)}
P_c\, e^{i \delta^P_c},
\nonumber\\
A_{10}\, e^{i \delta_{10}}
&=&
\frac{\sin{(\gamma - 5^\circ)}}{\sin{(5^\circ)}}
T_c\, e^{i \delta^T_c}
+
P_c\, e^{i \delta^P_c},
\label{reproduce-510-c}
\ea
Therefore,
in the $\{ 5^\circ, 10^\circ \}$ basis
\be
\lambda_{\pi^+ \pi^-}
=
e^{- 2 i \tilde{\beta}}\,
\frac{e^{- i 5^\circ} + z\, e^{- i 10^\circ}}{
e^{i 5^\circ} + z\, e^{i 10^\circ}}
=
e^{- 2 i (\tilde{\beta} - 5^\circ)}\,
\frac{1 + z\, e^{- i 5^\circ}}{1 + z\, e^{i 5^\circ}},
\label{lambda_in_510}
\ee
where
\be
z = \frac{A_{10}\, e^{i \delta_{10}}}{
A_{5}\, e^{i \delta_{5}}}
=
\frac{- \sin{(\gamma - 5^\circ)} + z_c \sin{(5^\circ)}
}{
\sin{(\gamma - 10^\circ)} - z_c \sin{(10^\circ)}
}.
\ee
Assuming that $\tilde{\beta}$ has been measured,
the two observables in $\lambda_{\pi^+ \pi^-}$ depend on only
two parameters:
the phase and magnitude of $z$.

For completeness,
we include the $p$-convention case,
where $\{ \phi_{A1}, \phi_{A2} \} = \{ \pi, - \beta \}$.
We find
\be
\lambda_{\pi^+ \pi^-}
=
e^{- 2 i \tilde{\beta}}\,
\frac{1 + z_p\, e^{i \beta}}{
1 + z_p\, e^{- i \beta}},
\label{lambda_in_p}
\ee
where
\ba
z_p
&=& e^{i \pi}\frac{M_{2p}\, e^{i \delta_{2p}}}{
M_{1p}\, e^{i \delta_{1p}}}
\nonumber\\
&=&
- \frac{\sin{\gamma}
}{
\sin{(\beta + \gamma)} + z_c \sin{\beta}
}
=
\frac{- \sin{\gamma}+ z_t \sin{\beta}
}{
\sin{(\beta + \gamma)}
},
\ea
which,
if the CKM is unitary, turns into
\be
z_p
=
- \frac{R_t}{1 + z_c R_b}
=
- R_t + z_t R_b\ .
\ee

To summarize,
let us assume that $\tilde{\beta}$ has been measured,
and that we wish to interpret the
two measurements contained in $\lambda_{\pi^+ \pi^-}$
(\textit{i.e.}, its magnitude and phase, or, alternatively,
$S_{\pi^+ \pi^-}$ and $C_{\pi^+ \pi^-}$).
We can fit these two observables in a variety of ways:
with two quantities ($z$),
in the $\{ 5^\circ, 10^\circ \}$ basis;
with three quantities ($z_c$ and $\gamma$),
in the $c$-convention;
or with four quantities ($z_t$, $\gamma$, and $\beta$),
in the $t$-convention.
As we have just explained,
there is no inconsistency,
despite the different parameter counting in each case.

\subsection{\label{subsec:pipi-isospin}The isospin
analysis in $B \rightarrow \pi \pi$}

In this section,
we extend the Gronau-London isospin analysis
\cite{GL},
allowing for the presence of new physics effects.
We face it not as a way to ``trap the penguin'' but,
rather,
as a way to determine
\be
\lambda_{\pi^+ \pi^0}
=
\frac{q}{p} \frac{A(B^- \rightarrow \pi^- \pi^0)}{
A(B^+ \rightarrow \pi^+ \pi^0)}.
\label{nonsense}
\ee
In principle,
the expression on the RHS of Eq.~(\ref{nonsense}) does not make sense;
it is no even rephasing invariant under independent
phase transformations of $B^0_d$,
$\overline{B^0_d}$, 
$B^+$,
and $B^-$.
However,
if one assumes isospin symmetry,
then $B^+$ must transform with $B^0_d$ and
$B^-$ with $\overline{B^0_d}$,
and Eq.~(\ref{nonsense}) becomes rephasing invariant.
Although
$\lambda_{\pi^+ \pi^0}$ is not measurable directly by any single
experiment,
since it involves both neutral and charged $B$ states,
we will now show that it can be determined experimentally through
the standard isospin analysis,
even in the presence new physics in both mixing and 
some type of new physics in decay (see below).

In the isospin analysis of Gronau and London one requires
the observables
$C_{+-}$, $S_{+-}$, $B_{+-}$,
$C_{+0}$, $B_{+0}$,
$C_{00}$, and $B_{00}$.
Here and henceforth,
the sub-indexes refer to the charges of the final state,
and we have used
\be
B_f = \frac{|A_f|^2 + |\bar A_f|^2}{2},
\label{B_f}
\ee
which is proportional to the untagged (charge averaged)
branching ratio into the final state $f$.
Eqs.~(\ref{C_f_detail}) and (\ref{B_f}) may be inverted,
\ba
|A_f|^2 &=& B_f (1 + C_f),
\nonumber\\
|\bar A_f|^2 &=& B_f (1 - C_f),
\ea
meaning that the magnitudes of the decay amplitudes are
determined from the observables $C_f$ and $B_f$
obtained in time-integrated decay rates.

Although we allow for new physics in the decay amplitudes,
we will assume that it obeys two properties:
i) that it does not produce large $\Delta I = 5/2$
contributions to the decay amplitudes;
and ii) that it does not produce large isospin-violating
contributions.
Under these conditions,
the decay amplitudes obey two triangle relations,
\ba
A_{+0} &=& \frac{1}{\sqrt{2}} A_{+-} + A_{00},
\nonumber\\
\bar A_{+0} &=& \frac{1}{\sqrt{2}} \bar A_{+-} + \bar A_{00}.
\label{isospin_triangle}
\ea
Since we know the magnitudes of all the decay amplitudes,
we may calculate
\ba
R & = &
\mbox{Re} \left( A_{+0} A_{+-}^\ast \right)
=
\frac{|A_{+0}|^2 + 1/2 |A_{+-}|^2 - |A_{00}|^2}{\sqrt{2}},
\nonumber\\
\bar R & = &
\mbox{Re} \left( \bar A_{+0} \bar A_{+-}^\ast \right)
=
\frac{|\bar A_{+0}|^2 + 1/2 |\bar A_{+-}|^2 - |\bar A_{00}|^2
}{\sqrt{2}},
\nonumber\\
\rho &=& |A_{+0} A_{+-}^\ast|,
\nonumber\\
\bar \rho &=& |\bar A_{+0} \bar A_{+-}^\ast|,
\ea
from which we can determine the complex vectors
\ba
A_{+0} A_{+-}^\ast &=& R \pm i \sqrt{\rho^2 - R^2},
\nonumber\\
\bar A_{+0} \bar A_{+-}^\ast &=& \bar R \pm i \sqrt{\bar{\rho}^2 - \bar R^2}.
\ea

Recall that the measurements of $C_{+-}$ and $B_{+-}$
yield $|A_{+-}|$ and $|\bar A_{+-}|$.
We continue to assume that $|q/p|=1$,
thus determining $|\lambda_{+-}|$.
Combining this with $S_{+-}$ determines also the
phase of $\lambda_{+-}$,
up to a two-fold discrete ambiguity,
\textit{c.f.\/} Eq.~(\ref{Lf_from_CS}).
Therefore,
$\lambda_{+-}$ is known experimentally.
Now we notice that
\be
\lambda_{+0} \lambda_{+-}^\ast
=
\left| \frac{q}{p} \right|^2
\frac{\bar A_{+0} \bar A_{+-}^\ast}{A_{+0} A_{+-}^\ast},
\ee
leading to
\be
\lambda_{+0} =
\frac{1}{\lambda_{+-}^\ast}
\frac{\bar R \pm i \sqrt{\bar{\rho}^2 - \bar R^2}}{
R \pm i \sqrt{\rho^2 - R^2}}.
\label{my_isospin}
\ee
Thus,
under the assumptions about the new physics mentioned above,
the isospin analysis allows us to determine the complex
quantity $\lambda_{+0}$ (up to an eight-fold discrete ambiguity),
which is the result we wanted to prove.

As a particular case,
we consider first how this analysis applies to the SM.
There,
the $\Delta I = 5/2$ contributions \cite{5/2},
the electroweak penguins \cite{electroweak}
and other isospin violating contributions to this channel
are small \cite{isospin_violating}.
Neglecting them,
the isospin triangle relations in Eqs.~(\ref{isospin_triangle})
remain.
Also,
the penguin diagrams only contribute to the
$\Delta I = 1/2$ term which, moreover,
does not contribute to $A_{+0}$.
As a result,
$A_{+0}$ has the phase $\gamma$ of the tree level diagram
and the ``reconstructed observable'' is
\be
\lambda_{\pi^+ \pi^0} = e^{2 i \alpha},
\ee
where $\alpha = \pi - \beta - \gamma$.
Thus, in the SM,
one predicts that $|\lambda_{+0}|=1$ (\textit{i.e.,} $C_{+0}=0$),
and that the phase reconstructed from the measurable quantities
through Eq.~(\ref{my_isospin}) is the CKM phase $\alpha$. 
(We may allow for models in which the
new physics contributes exclusively 
to a new phase in the mixing by substituting $\alpha$ by
$\tilde{\alpha} = \pi - \tilde{\beta} - \gamma$.)
In the SM, this does not coincide with the
phase measured in $B_d \rightarrow \pi^+ \pi^-$ decays,
\be
\lambda_{\pi^+ \pi^-} = |\lambda_{\pi^+ \pi^-}| e^{2 i \alpha_{\rm eff}},
\ee
and Eq.~(\ref{my_isospin}) reads
\be
e^{2 i \alpha} = e^{2 i \alpha_{\rm eff}} e^{- 2 i \delta_\alpha}.
\ee
The factor of two multiplying $\alpha$ in the exponent leads to a further
two-fold ambiguity,
meaning that this isospin analysis determines $\alpha$ with
a sixteen-fold discrete ambiguity.
It is well known that the SM isospin construction may be
used in order to place bounds on $\delta_{\alpha}$
(and, thus, on how much the $\alpha_{\rm eff}$ measured in
$B_d \rightarrow \pi^+ \pi^-$ decays differs from
the CKM phase $\alpha$) even if the observable $C_{00}$
is not known to the required precision \cite{GQ}.

We should also mention that
there is a relation between $\lambda_{+0}$ and
$\lambda_{00}$ similar to the one between
$\lambda_{+0}$ and $\lambda_{+-}$ in Eq.~(\ref{my_isospin}).
This means that the observables
$B_{+-}$, $C_{+-}$, $S_{+-}$, $B_{00}$, $C_{00}$,
$B_{+0}$, and $C_{+0}$ determine not only
$\lambda_{+0}$ but also $S_{00}$.
If the measurement of $S_{00}$ were feasible,
we would have a cross-check on the isospin construction,
as well as a reduction in the ambiguity in the
determination of $\alpha$ to four-fold.

In the SM,
Eq.~(\ref{my_isospin}) is just a new way of interpreting
the Gronau-London isospin analysis.
However,
Eq.~(\ref{my_isospin}) tells us what portions of the isospin analysis
remain in more general models;
namely, the determination of $\lambda_{+0}$.

Let us now turn to the interpretation of this general isospin analysis,
in the light of reparametrization invariance.
If $C_{+0}=0$,
then we know from subsection~\ref{subsec:experimental} that $\lambda_{+0}$
determines one ``experimental'' weak phase.
In theories in which there is really only one weak phase
mediating the decay,
then $\lambda_{+0}$ measures that weak phase.
But,
as mentioned in subsection~\ref{subsubsec:theory1},
$C_{+0}=0$ might be due to a negligible strong phase difference between two
amplitudes with different weak phases.
In either case,
the SM is excluded if the weak phase of $\lambda_{+0}$ differs
from the SM $2 \alpha$ (up to electroweak penguins).
Of course,
the SM is trivially excluded if $C_{+0}$
is sufficiently different from zero.

Let us now imagine that $\lambda_{+0}$ has been determined,
has unit magnitude and that its phase agrees (within errors)
with the SM $2 \alpha$.
Does this guarantee that the SM is the correct description
of the decay?
We will return to this question at the end of section~\ref{sec:single}.

\section{\label{sec:psiK}The decay $B \rightarrow \psi K$}

In the SM,
the $\bar b \rightarrow \bar s$ decays leading to
$B \rightarrow \psi K$ can get contributions
proportional to any of the CKM structures in
Eq.~(\ref{decay}):
\ba
V_{ub}^\ast V_{us}
&\sim&
A \lambda^4 R_b e^{i (\gamma + \chi^\prime)},
\label{btos_phase_1}
\\
V_{cb}^\ast V_{cs}
&\sim&
A \lambda^2,
\label{btos_phase_2}
\\
V_{tb}^\ast V_{ts} 
&\sim&
- A \lambda^2 R_t e^{i \chi}.
\label{btos_phase_3}
\ea
The quantities appearing on the RHS are defined in the appendix
in a rephasing invariant way,
but a convenient phase convention has been used in equating the
RHS to the LHS.
The phases already reflect the structure of a possible non-unitary
CKM matrix,
but the magnitudes reflect the SM (unitarity) constraints.
Substitution of Eqs.~(\ref{btos_phase_1})-(\ref{btos_phase_3})
in Eq.~(\ref{unitarity})
leads to 
\be
- \lambda^2 R_b e^{i (\gamma + \chi^\prime)} + e^{i \chi} = 1,
\label{squashed_unit_triangle}
\ee
which represents a very ``squashed'' unitarity triangle.
Thus, 
if inspired by the weak phases which appear in the SM description
of $B \rightarrow \psi K$ decays,
we are lead to choosing two among the weak phases
$\gamma + \chi^\prime$, $0$, and $\chi+\pi$.
The equivalent to the $c$-convention in
Eq.~(\ref{A_pipi_c}) is
\be
A_{\psi K}
=
M_{1c}\, e^{i \delta_{1c}} e^{i (\gamma + \chi^\prime)}
+
M_{2c}\, e^{i \delta_{2c}};
\label{A_psiK_c}
\ee
the equivalent to the $t$-convention in
Eq.~(\ref{A_pipi_t}) is
\be
A_{\psi K}
=
M_{1t}\, e^{i \delta_{1t}} e^{i (\gamma+ \chi^\prime)}
+
M_{2t}\, e^{i \delta_{2t}} e^{i (\chi + \pi)};
\label{A_psiK_t}
\ee
and the equivalent to the $p$-convention is
\be
A_{\psi K}
=
M_{1p}\, e^{i \delta_{1p}}
+
M_{2p}\, e^{i \delta_{2p}} e^{i (\chi + \pi)}.
\label{A_psiK_p}
\ee
But we may equally well use any other two weak phases as our basis.
The relation between the several options follows
the analysis in sections~\ref{subsec:decay_amp_pipi} and
\ref{subsec:neutral_pipi}.

In the SM,
the $\bar b \rightarrow \bar s$ decays are predicted to
depend (almost) on a single weak phase.
A rough argument is usually presented based on the
CKM structures in Eqs.~(\ref{btos_phase_1})--(\ref{btos_phase_3}),
and can be seen clearly in Eq.~(\ref{A_psiK_p}),
where the second phase, $\chi \sim \lambda^2$,
is very close to the first phase, $0$.
To revisit this argument in the other two conventions we must
notice that,
although the phase difference is large,
the SM predicts a hierarchy in the magnitudes of those
cases.
Indeed,
what matters for the difference of $\lambda_f$ from
the ideal case in which it measures a pure weak phase 
($\phi_2 = \phi_{A2}- \phi_M$) 
is the product \cite{done}
\be
\frac{M_1}{M_2} \sin{(\phi_{A1} - \phi_{A2})}.
\ee
Table~\ref{products} shows the hierarchy of the two magnitudes
imposed by the magnitudes of the CKM matrix elements
and the phase difference in the three conventions.
\begin{table}
\caption{\label{products}CKM predictions for the (CKM part of the)
ratio of magnitudes and for the difference of weak phases between
the two terms,
in the various conventions.}
\begin{ruledtabular}
\begin{tabular}{lccc}
convention & $\left(M_1/M_2\right)_{\rm CKM\  part}$
                   & $\sin \left( \phi_{A1} - \phi_{A2} \right)$ & product
\\
\hline
$c$-type & $\sim R_b \lambda^2$
                   & $\sin \left( \gamma + \chi^\prime  \right) \sim 1$
                                             & $\sim \lambda^2$
\\
$t$-type & $\sim R_b \lambda^2$
             & $\sin \left( \gamma + \chi^\prime -\chi - \pi \right) \sim 1$ 
                                             & $\sim \lambda^2$
\\
$p$-type & $\sim 1$
                   & $\sin \left( -\chi - \pi \right) \sim \lambda^2$ 
                                             & $\sim \lambda^2$
\\
\end{tabular}
\end{ruledtabular}
\end{table}
%
The product is always small,
leading to the conclusion that,
barring a large compensating hierarchy arising from the
hadronic matrix elements,
the SM predicts that $\lambda_{\psi K_S}$ does measure a single weak phase.
Several calculations of the hadronic matrix elements involved
show that these do not offset the hierarchy seen
in Table~\ref{products}.
As a result,
using the $c$--type convention within the SM,
one finds
\ba
\lambda_{\psi K_S} &=&
- e^{- 2 i (\tilde{\beta} + \chi^\prime)}
\frac{1+|M_{1c}/M_{2c}|
e^{-i(\gamma + \chi^\prime) e^{i (\delta_{1c} - \delta_{2c})}} 
}{
1+|M_{1c}/M_{2c}|
e^{i(\gamma + \chi^\prime) e^{i (\delta_{1c} - \delta_{2c})}} 
}
\\
&\approx&
- e^{- 2 i (\tilde{\beta} + \chi^\prime)},
\ea
where $|M_{1c}/M_{2c}| \ll 1$ was used in obtaining the second line.
(We mention in passing that the decay $B_d \rightarrow \phi K_S$
may be written in a similar fashion.
However,
there is some experimental evidence that, in that case,
$|M_{1c}/M_{2c}|$ may not be much smaller than unity,
meaning that the presence of a second weak phase must be
taken into account.)

The fact that $\lambda_{\psi K_S}$ is given by a single
weak phase and that that phase agrees with the SM
expectation ($2 \beta$) is confirmed experimentally to very high accuracy
\cite{HFAG}:
\ba
|\lambda_{B_d \rightarrow (c \bar c) K}|
&=&
0.969 \pm 0.028\ ,
\nonumber\\
\sin{\left(
\arg{\lambda_{B_d \rightarrow (c \bar c) K}}
\right)}
&=&
0.725 \pm 0.037\ .
\ea
Since $\lambda_{+0}$ and $\lambda_{\psi K_S}$ depend on a
single weak phase in the SM,
we now turn to a detailed analysis of those situations,
in the light of reparametrization invariance.

\section{\label{sec:single}Decays which depend on a single weak phase}

\subsection{\label{subsec:versus}Reparametrization invariance versus
rephasing invariance}

Before we proceed,
we should clarify a few questions.
First: so far, we have been using the weak phases in decay,
$\phi_{A1}$ and $\phi_{A2}$,
and the weak phase in mixing,
$\phi_M$.
However,
these quantities are not invariant under a rephasing of the
meson states or of the quark field operators \cite{BLS2}.
The quantities which are rephasing invariant are
$\phi_1 = \phi_{A1} - \phi_M$ and $\phi_2 = \phi_{A2} - \phi_M$,
which show up in Eq.~(\ref{Lf_r_and_delta}).
A similar argument applies to $\delta = \delta_2 - \delta_1$.
For this reason, it is sometimes
better to combine Eqs.~(\ref{A_f}),
(\ref{Abar_fbar}),
and (\ref{q/p_parametrization}) into
\ba
\sqrt{\frac{q_B}{p_B}}
\bar A_f
&=&
\eta_f\, e^{i \delta_1}\, 
\left( M_1 e^{- i \phi_{1}} + M_2 e^{- i \phi_{2}} e^{i \delta} \right),
\label{Abar_fbar_rephasing_invariant}
\\
\sqrt{\frac{p_B}{q_B}}
A_f
&=&
e^{i \delta_1}
\left( M_1 e^{i \phi_{1}} + M_2 e^{i \phi_{2}}  e^{i \delta} \right),
\label{A_f_rephasing_invariant}
\ea
which ($\delta_1$ aside) refer only to rephasing invariant quantities.

Second: this rephasing invariance is \textit{not} the same as the
reparametrization invariance we discuss in this article.
To see this, we consider some decay amplitude with phases
$\phi_{A1}$ and $\phi_{A2}$. It is true that, through rephasings,
we may change these phases into $\phi_{A1} - \phi^\prime$ and
$\phi_{A2} - \phi^\prime$,
respectively.
However,
the difference between the two phases remains fixed.
This is \textit{not} the case when we consider a general
reparametrization from the phases $\{\phi_{A1}, \phi_{A2}\}$
into some other basis $\{\phi_{A1}^\prime, \phi_{A2}^\prime \}$,
such that
$\phi_{A2}^\prime - \phi_{A1}^\prime \neq \phi_{A2} - \phi_{A1}$.
Section~\ref{subsec:decay_amp_pipi} provides
ample illustrations of this point.

Third: under reparametrization of the decay amplitudes,
the rephasing invariant phases $\phi_1$ and $\phi_2$ are
changed into $\phi_1^\prime = \phi_{A1}^\prime - \phi_M$
and $\phi_2^\prime = \phi_{A2}^\prime - \phi_M$,
respectively.
As we stressed when discussing the phases
in the decay amplitudes,
this means that,
in general,
these phases are not measurable in a single decay.
What we can measure is the phase of $\lambda_f$,
which,
as we showed in section~\ref{subsec:experimental},
measures a single weak phase if and only if $C_f=0$.

\subsection{\label{subsec:single_weak}Interpreting results
which depend on a single weak phase}

Imagine that $C_f=0$ and that $S_f$ has been measured.
Then,
as we showed in section~\ref{subsec:experimental},
$\lambda_f$ depends on a single ``experimental'' weak phase
$\phi_3 = \phi_{A3} - \phi_M$,
\be
\lambda_f = \eta_f e^{- 2 i \phi_3}.
\label{experiment_3}
\ee
We may now ask what type of theories reproduce this result.
Clearly,
a theory in which the decay is given by a single weak phase
and has
\be
e^{- i \delta_3} \,\sqrt{\frac{q_B}{p_B}}
\bar A_f
=
\eta_f\,  M_3\, 
e^{- i \phi_3}
\ee
reproduces this result.
(In the usual parlance of phases in mixing, $\phi_M$,
and phases in decay, $\phi_{A3}$,
this may be obtained with different choices for these phases,
$\phi_{A3}^\prime$ and $\phi_M^\prime$,
as long as
$\phi_3 = \phi_{A3} - \phi_M = \phi_{A3}^\prime - \phi_M^\prime$.)

Can this result be obtained in a different way with the
parametrization in Eqs.~(\ref{Abar_fbar_rephasing_invariant})
and (\ref{A_f_rephasing_invariant})?
Eq.~(\ref{sol_phi3}) answers this question.
It shows that a theory with two distinct weak phases
can reproduce the ``experimental'' result of Eq.~(\ref{experiment_3})
as long as the two diagrams do not have a strong phase difference
and their weak phases obey Eq.~(\ref{sol_phi3}).

We will now address a more subtle question.
Let us assume that one knows that $\lambda_f$ is given by a
single weak phase within the SM,
as in Eq.~(\ref{experiment_3}).
This occurs,
for instances,
in $\lambda_{+0}$ discussed at the end of
section~\ref{subsec:pipi-isospin},
or in $\lambda_{\psi K_S}$ discussed in
section~\ref{sec:psiK}.
In each case,
we know what the SM diagrams are and that,
when there are several,
they share (at least very approximately)
their weak phase.
We also assume that the experiments confirm that this
is indeed given by a single weak phase
($C_f = 0$)
and that this phase coincides with the SM prediction
($S_f$ has been measured and it agrees with the SM expectation).
Now we entertain the possibility that
there is only one new physics diagram
contributing to this decay,
with a new weak phase $\phi_{A4}$,
and that there is no new contribution to the mixing.
Then $\phi_3$ has the SM value and there is now a
new weak phase $\phi_4 = \phi_{A4} - \phi_M \neq \phi_3$.
Clearly,
because we wish to reproduce $\lambda_f$,
the strong phase difference between the two diagrams
must vanish.
But then,
the new weak phase must be such that
\be
e^{- 2 i \phi_3}
=
\frac{M_3 e^{- i \phi_3} + M_4 e^{- i \phi_4}}{
M_3 e^{i \phi_3} + M_4 e^{i \phi_4}}
=
\frac{ 1 + r e^{i (\phi_3 - \phi_4)}}{
1 + r e^{- i (\phi_3 - \phi_4)}}\, 
e^{- 2 i \phi_3}.
\label{incredible}
\ee
This equation is only possible if $\phi_4 = \phi_3 + n \pi$,
contradicting our hypothesis.
This means that in experiments which:
\begin{itemize}
\item are predicted in the SM to depend on a single weak phase;
\item the ``experimental'' weak phase has been measured
(\textit{i.e.}, $C_{f}=0$ and $S_f$ has been measured);
\item and the weak phase measured coincides with that predicted
by the SM;
\end{itemize}
there is absolutely no possibility that the new physics
brings a new weak phase exclusively to the decay
which is different from the SM one
(even if the strong phase difference between the two diagrams
vanishes).
It is easy to see that having more than one new physics diagram
does not help \cite{not_easy}.

The only way out of this conclusion occurs if the new physics
contributes with a new weak phase to the decay and also with
a new weak phase to the mixing.
In that case,
the full theory has a new mixing phase $\phi_M^\prime$,
leading to two weak phases $\phi_3^\prime \neq \phi_3$
and $\phi_4^\prime$ and Eq.~(\ref{incredible})
becomes
\be
e^{- 2 i \phi_3}
=
\frac{M_3 e^{- i \phi_3^\prime} + M_4 e^{- i \phi_4^\prime}}{
M_3 e^{i \phi_3^\prime} + M_4 e^{i \phi_4^\prime}}
=
\frac{ 1 + r e^{i (\phi_3^\prime - \phi_4^\prime)}}{
1 + r e^{- i (\phi_3^\prime - \phi_4^\prime)}}\, 
e^{- 2 i \phi_3^\prime},
\label{incredible_2}
\ee
which might have nontrivial solutions.

\section{\label{sec:conclusions}Conclusions}

In this article,
we point out that a generic $B \rightarrow f$ decay amplitude
may be written as a sum of two terms,
corresponding to a pair of weak phases $\{ \phi_{A1}, \phi_{A2} \}$
chosen completely at will
(as long as they do not differ by a multiple of $180^\circ$).
Clearly,
physical observables may not depend on this choice;
we designate this property by ``reparametrization invariance''.

We explore some of the unusual features of reparametrization invariance.
For example,
we show that the relations between the $c$-convention and $t$-convention
in $B \rightarrow \pi \pi$ decays have nothing to do with CKM unitarity
and that Eqs.~(\ref{reproduce}) would hold even
if ``$\beta$'' and ``$\gamma$''
were merely names for some weak phases with no connection to the SM
whatsoever.
This allows us to explain the apparent discrepancy in the
parameter counting of $\lambda_{\pi^+ \pi^-}$ for some specific
choices for the weak phases utilized as a basis,
\textit{c.f.\/} Eqs.~(\ref{lambda_in_c-convention}),
(\ref{lambda_in_t-convention}),
and (\ref{lambda_in_510}).
We have extended the isospin analysis of Gronau and London to cases
with new physics in the mixing and (some types of) new physics
in the decay amplitudes,
viewing it as a way to measure $\lambda_{\pi^+ \pi^0}$. 

As a result of reparametrization invariance,
the weak phases used as a basis cannot (in general) be measured
in any single decay.
To study the one exception,
we showed that $C_f = 0$ if and only if the decay amplitude
is dominated by a single weak phase.
Thus,
decays in which $C_f=0$ do determine one ``experimental'' weak phase.
Nevertheless,
this can only be turned into knowledge about a SM weak phase
in those cases in which the SM description of the decay
depends on only one weak phase.
Finally,
in experiments with $C_f=0$,
which are predicted in the SM to depend on a single weak phase
and do indeed reproduce the phase expected
(known from other constraints on the unitary CKM matrix)
there is absolutely no possibility that the new physics
brings a new weak phase exclusively to the decay
which is different from the SM one
(even if the strong phase difference between the two diagrams
vanishes).
This ``theorem'' might only be evaded if the new physics
contributes with a new weak phase to the decay and also with
a new weak phase to the mixing.

\begin{acknowledgments}
We are grateful to L.\ Wolfenstein for discussions.
This work was supported by the Portuguese \textit{Funda\c{c}\~{a}o para
a Ci\^{e}ncia e a Tecnologia} (FCT) under the contract CFTP-Plurianual (777).
In addition, F.\ J.\ B.\ is partially supported 
by the spanish M.\ E.\ C.\ under FPA2002-00612 and HP2003-0079
(``Accion Integrada hispano-portuguesa''),
and J.\ P.\ S.\ is supported in part by project POCTI/37449/FNU/2001,
approved by the Portuguese FCT and POCTI,
and co-funded by FEDER. 
\end{acknowledgments}

\appendix*
\section{\label{sec:notation}Notation}

\subsection{Observables in decay rates}

The time-dependent decay rate of a neutral meson into the final state $f$
may be written as
\ba
\Gamma [ B^0 (t) \ra f ]
& = &
\frac{\left| A_f \right|^2 + \left| \bar A_f \right|^2}{2}\ 
e^{- \Gamma\, t}\,
\left\{
\cosh{\left( \frac{\Delta \Gamma\, t}{2} \right)}
+
D_f
\sinh{\left( \frac{\Delta \Gamma\, t}{2} \right)}
\right.
\nonumber\\*[2mm]
& &
\hspace{3.5cm}
\left.
+\;
C_f
\cos{(\Delta m\, t)}
-
S_f
\sin{(\Delta m\, t)}
\right\},
\nonumber\\*[4mm]
\Gamma [ \overline{B^0} (t) \ra f ]
& = &
\frac{\left| A_f \right|^2 + \left| \bar A_f \right|^2}{2}\ 
e^{- \Gamma\, t}\,
\left\{
\cosh{\left( \frac{\Delta \Gamma\, t}{2} \right)}
+
D_f
\sinh{\left( \frac{\Delta \Gamma\, t}{2} \right)}
\right.
\nonumber\\*[2mm]
& &
\hspace{3.5cm}
\left.
-\;
C_f
\cos{(\Delta m\, t)}
+
S_f
\sin{(\Delta m\, t)}
\right\},
\label{master_Bs}
\ea
where
\ba
D_f & \equiv &
\frac{2 \mbox{Re}( \lambda_f )}{1 + |\lambda_f|^2}
\label{D_f}
\\
C_f & \equiv &
\frac{1 - |\lambda_f|^2}{1 + |\lambda_f|^2}
\label{C_f}
\\
S_f & \equiv &
\frac{2 \mbox{Im}( \lambda_f )}{1 + |\lambda_f|^2}.
\label{S_f}
\ea
Therefore,
\be
\lambda_f =
\frac{1}{1+C_f} \left( D_f + i S_f \right)
\label{Lf_from_DCS}
\ee
is a physical observable, and
\be
D_f^2 + C_f^2 + S_f^2 = 1.
\label{D_C_S}
\ee
If the width difference is too small (as it happens in the
$B_d$ system),
then we can set $\Delta \Gamma = 0$ and
$D_f$ is not measured.
It can be inferred from 
Eq.~(\ref{D_C_S}) with a twofold ambiguity,
meaning that $\lambda_f$ is determined from
Eq.~(\ref{Lf_from_DCS}) with that twofold ambiguity.

\subsection{Phase structure of a generalized CKM matrix}

It is interesting to note that,
even if one goes beyond the SM,
there are only four irremovable phases in the 
generalized CKM matrix \cite{BLS},
which we may chose to be
\ba
\beta
&\equiv&
\arg{\left( - \frac{V_{cd} V_{cb}^\ast}{V_{td} V_{tb}^\ast} \right)},
\label{beta}
\\
\gamma
&\equiv&
\arg{\left( - \frac{V_{ud} V_{ub}^\ast}{V_{cd} V_{cb}^\ast} \right)},
\label{gamma}
\\
\chi &\equiv&
\arg{\left( - \frac{V_{cb} V_{cs}^\ast}{V_{tb} V_{ts}^\ast} \right)},
\label{chi}
\\
\chi^\prime &\equiv&
\arg{\left( - \frac{V_{us} V_{ud}^\ast}{V_{cs} V_{cd}^\ast} \right)}.
\label{chi_prime}
\ea
Such a generalized CKM matrix will cease to be unitary,
but we may parametrize its phase structure as \cite{BLS}
\be
\arg V =
\left(
\begin{array}{ccc}
	0 & \chi^\prime & - \gamma\\
	\pi & 0 & 0\\
	- \beta& \pi + \chi & 0 
\end{array}
\right),
\label{arg_V}
\ee
where a convenient phase convention has been chosen.
The phase $\alpha$ is \textit{defined} by
$\alpha = \pi - \beta - \gamma$.
In the SM,
these phases are related to the Wolfenstein parameters through
\ba
R_t e^{- i \beta} &\approx& 1 - \rho - i \eta,
\label{R_t_Wolf}
\\
R_b e^{- i \gamma} &\approx& \rho - i \eta,
\label{R_b_Wolf}
\\
\chi &\approx& \lambda^2 \eta,
\label{chi_Wolf}
\\
\chi^\prime &\approx& A^2 \lambda^4 \eta,
\label{chi_prime_Wolf}
\ea
where
\ba
R_b &=& \left| \frac{V_{ud} V_{ub}}{V_{cd} V_{cb}} \right| ,
\label{R_b}
\\
R_t &=& \left| \frac{V_{td} V_{tb}}{V_{cd} V_{cb}} \right| .
\label{R_t}
\ea
This allows us to use a schematic form for the SM CKM matrix which
keeps this phase structure but only expands each magnitude
to leading order in $\lambda$.
This is not a consistent expansion,
but allows us to see where the phases would come in if we expanded
everything up to a sufficient high power of $\lambda$ \cite{Prague04}:
\be
V \approx
\left(
\begin{array}{ccc}
1 
&
	\lambda e^{i \chi^\prime}&
	A \lambda^3 R_b e^{- i \gamma}
\\*[2mm]
- \lambda &
	1 
&
	A \lambda^2
\\*[2mm]
A \lambda^3 R_t e^{- i \beta} &
	- A \lambda^2 e^{i \chi}&
	1
\end{array}
\right).
\label{crazy_V}
\ee
The ranges for $\chi$ and $\chi^\prime$ are discussed in
reference \cite{chi} for some models of new physics.


\begin{thebibliography}{99}
\bibitem{details}
Notice that,
in equating the first line of Eq.~(\ref{two_conventions}) to its second line,
a specific phase choice has been made for the relative phase
between the two conventions.
Thus,
the strong phases in the two conventions become tied together.
For example,
one may no longer choose the phases $\delta_1$ and $\delta^\prime_1$
independently.
This can be seen clearly in Eqs.~(\ref{relate}),
where setting $\delta_1=\delta^\prime_1=0$ would lead
to the crazy conclusion that $\delta_2=\delta^\prime_2=0$,
thus precluding direct CP violation.
\bibitem{gauge}
To be exact, for the case of penguin diagrams,
it is only the differences appearing in Eqs.~(\ref{uc})-(\ref{ct})
which are gauge invariant.
\bibitem{pi} 
There are also the possibilities that
$r= \infty$,
or $\phi_1 - \phi_2 = \pi$,
or $\delta = \pi$,
whose analysis follow a similar pattern.
\bibitem{relate_notations}
The parameters used in Eqs.~(\ref{reproduce})-(\ref{reproduce-p-t})
are related to those in Eqs.~(\ref{uc})-(\ref{ct})
by:
%
\begin{eqnarray}
T_c e^{i \delta^T_c}
e^{i \gamma}
&=&
V_{ub}^\ast V_{ud} \left( A_u - A_t \right),
\nonumber\\
P_c e^{i \delta^P_c}
e^{i \pi}
&=&
V_{cb}^\ast V_{cd} \left( A_c - A_t \right),
\nonumber\\
T_t e^{i \delta^T_t}
e^{i \gamma}
&=&
V_{ub}^\ast V_{ud} \left( A_u - A_c \right),
\nonumber\\
P_t e^{i \delta^P_t}
e^{- i \beta}
&=&
V_{tb}^\ast V_{td} \left( A_t - A_c \right),
\nonumber\\
M_{1p} e^{i \delta_{1p}}
e^{i \pi}
&=&
V_{cb}^\ast V_{cd} \left( A_c - A_u \right),
\nonumber\\
M_{2p} e^{i \delta_{2p}}
e^{- i \beta}
&=&
V_{tb}^\ast V_{td} \left( A_t - A_u \right).
\end{eqnarray}
\bibitem{not_easy}
Imagine some decay having a SM contribution 
with magnitude $\tilde M_3$,
weak phase $\phi_3$,
and strong phase $\tilde{\delta}_3$.
Now imagine that there are several new physics contributions,
with different magnitudes $\tilde M_k$,
weak phases $\phi_k$,
and strong phases $\tilde \delta_k$ ($k = 4, 5, \dots$),
%
\ba
\sqrt{\frac{q_B}{p_B}}
\bar A_f
&=&
\eta_f\, \sum_k \tilde M_k\, e^{- i \phi_k} e^{i \tilde{\delta}_k},
\nonumber\\
\sqrt{\frac{p_B}{q_B}}
A_f
&=&
\sum_k \tilde M_k\, e^{i \phi_k} e^{i \tilde{\delta}_k}.
\ea
%
Using a reasoning similar to the one in
Eqs.~(\ref{master})--(\ref{ak_bk}),
we may write everything in terms of two weak phases,
which we choose as the SM phase $\phi_3$ and the phase
$\phi_4$ of the first new physics diagram.
This reduces the general problem to the
simple case in Eq.~(\ref{incredible}).
\bibitem{done}
How the several conventions impact the argument for
the dominance of a single weak phase in $\bar b \rightarrow \bar s$ decays, 
has been discussed on pages 404--405 of \cite{BLS}.
\end{thebibliography}
\end{document}